# High temperature equilibrium of 3D and 2D chalcogenide perovskites.

*Prakriti Kayastha, Devendra Tiwari, Adam Holland, Oliver S. Hutter, Ken Durose, Lucy D. Whalley\* and Giulia Longo\**

Prakriti Kayastha, Devendra Tiwari, Oliver S. Hutter, Lucy Whalley, Giulia Longo

Department of Mathematics, Physics and Electrical Engineering, Northumbria University,

Ellison Place, Newcastle upon Tyne NE1 8ST, United Kingdom

Devendra Tiwari

School of Chemistry, University of Bristol, Bristol BS8 1TS, United Kingdom

Adam Holland

HORIBA UK Limited, Kyoto Close, Moulton Park, Northampton NN3 6FL, United Kingdom

Ken Durose

Department of Physics, Stephenson Institute for Renewable Energy, Chadwick Building, Peach St, Liverpool L23 9SD, United Kingdom

**Abstract**

Chalcogenide perovskites have been recently under the researchers' spotlight as novel absorber materials for photovoltaic applications. BaZrS$_3$, the most investigated compound of this family, shows a high absorption coefficient, a bandgap of around 1.8 eV, and excellent environmental and thermal stability. In addition to the 3D perovskite BaZrS$_3$, the Ba-Zr-S compositional space contains various 2-D Ruddlesden-Popper phases Ba$_{n+1}$Zr$_n$S$_{3n+1}$ (with n = 1, 2, 3) which have recently been reported. In this work it will be shown that at high temperature the Gibbs free energies of 3D and 2D perovskites are very close, suggesting that 2D phases can be easily formed at high temperatures. The analysis of the product of the BaS and ZrS$_2$ solid-state reaction, in different stoichiometric conditions, present a mixture of BaZrS$_3$ and Ba$_4$Zr$_3$S$_{10}$. To carefully resolve the composition, XRD, SEM and EDS analysis were complemented with Raman spectroscopy. For this purpose, the phonon modes, and the consequent Raman spectra, were calculated for the 3D and 2D chalcogenide perovskites, as well as for the binary precursors. This thorough characterization demonstrates the thermodynamic limitations and experimental difficulties in forming phase-pure chalcogenide perovskites through solid state synthesis, and the importance of using multiple techniques to soundly resolve the composition of these chalcogenide materials.

**Introduction**

Chalcogenide perovskites (CPVK) have recently been proposed as possible non-toxic alternatives to lead-based perovskites for photovoltaic applications thanks to their promising optoelectronic characteristics, including a bandgap that is suited for tandem solar cell applications.[1–4] Chalcogenide perovskites follow the formula $ABX_3$ with A, B and X representing an alkaline earth cation (2+), a transition metal cation (4+) and a chalcogenide anion (2-) respectively.[5–7] The most studied compound in this family is $BaZrS_3$, but other compositions have been reported.[8,9]

Accordingly, chalcogenide perovskites have been under recent scrutiny to evaluate and validate a variety of optoelectronic parameters. Reports suggest the possibility of tunable bandgaps through compositional engineering with high absorption coefficients in the visible range,[10] low effective masses[11] and, resultantly, high carrier mobilities.[12] Moreover, these materials should present improved thermal and chemical stability compared with the hybrid halide perovskites, being resistant towards high temperatures and atmospheric conditions.[13]

Notwithstanding these promising features, the development of CPVK-based devices is still hindered by the synthetic procedure necessary to prepare these perovskites. In practice, these materials need very high temperatures to be crystallized in the desired phase. For example, $BaZrS_3$ is produced by the solid-state reaction of the elemental or binary precursors at 800-1100°C for several hours or days. This requirement for high temperatures is limiting as it does not easily allow for thin film processing and device integration, and can lead to chalcogen loss via volatile precursors.

Understanding reaction mechanisms can help design fabrication procedures potentially having lower temperatures, and recent research has started to explore which processes may allow for this. For example, it has been proposed that favoring the formation of $BaS_3$, which has a lower melting point than BaS, can trigger the formation of $BaZrS_3$ at temperatures as low as 600°C.[14] Similarly, it has been reported that an excess of sulphur in the reaction mixture can significantly reduce the reaction time and temperature.[15]

The tuning of the precursors' stoichiometry can provide a path to find milder synthetic conditions for $BaZrS_3$ and other chalcogenide perovskites, but it can also lead to the formation of other unwanted perovskite phases. The ideal perovskite structure is formed from a three-dimensional network of corner sharing octahedra. However, there are other closely related perovskite-like phases that can be formed. For example, low dimensional Ruddlesden-Popper (RP) phases are known for chalcogenide perovskites, similarly to the oxide and halide perovskites. $Ba_{n+1}Zr_nS_{3n+1}$ for n = 1, 2, 3, (with the presence of low- and high-temperature polymorphs for $Ba_3Zr_2S_7$) have been reported in addition to the 3D structure $BaZrS_3$ (in which n = ∞).[6,13,16–18] For the corresponding perovskites with Hf ($Ba_{n+1}Hf_nS_{3n+1}$), additional Ruddlesden Popper phases with n = 4, 5 have been reported,[19] while no evidence is present for n > 3 in the Zr series. In these low dimensional phases, layers of 3D CPVK are alternated with layers of BaS, as represented in Figure 1. Interestingly, the bandgaps of these chalcogenide RP perovskites decrease as n increases, contrary to the oxide and halide counterparts, where the bandgap widens as n increases.[20–22] The different crystalline structure not only affects the bandgap of the perovskite, but can also affect other functional features, such as carrier transport and thermal or chemical stability. A deep understanding of the synthetic reaction, especially when non-stoichiometric conditions are explored, will be essential to control the formation of competing phases with distinct properties.

The most widely used technique for the assessment of different phases is X-ray diffraction (XRD). However, due to the structural similarity of the 3D and the 2D perovskites, the diffractogram peaks overlap, which make differentiating between species challenging. For example, in the case of samples formed by the mixture of $BaZrS_3$ and $Ba_3Zr_2S_7$, long data collection times and rigorous refinement (supported by compositional techniques such as Energy Dispersive X-ray Spectroscopy, EDS) are necessary to give a quantitative estimation of their compositions. As such, even with diffractograms presenting good angular resolution, intensity, and angular range, ambiguous assignation can still occur. It follows that good practice, especially for this family of materials, would be to combine crystallographic analysis with other complementary techniques that may also provide a clearer distinction between species. Vibrational spectroscopy is a good candidate technique to carry out this role, as it probes the local structure of a material, in contrast to XRD which probes the bulk response. In this letter it will be shown that the 3D and 2D perovskite structures each have a unique vibrational fingerprint that better distinguishes between materials in the Ba-Zr-S system and that Raman spectroscopy is therefore well suited for checking the phase purity of the compounds.

In this study XRD and Raman spectroscopy are combined to assess the main composition of the product resulting from the solid-state reaction of BaS and $ZrS_2$ at various ratios. To aid in our analysis, a first-principles thermodynamic model is used to demonstrate that at high temperatures the Gibbs free energy of the 3D and 2D CPVK materials are only a few kJ/mol apart, indicating that both can be formed during high temperature synthesis (>1000K). In addition, the Raman spectra for all known binary and ternary compositions in the Ba-Zr-S phase systems are calculated from first-principles. Despite the use of excess $ZrS_2$ in the reaction mixtures, all the resulting powders show deficiency of Zr and S, and present mixtures of $BaZrS_3$ and the Ruddlesden-Popper phase $Ba_4Zr_3S_{10}$. It will also be shown that peaks' assignation can be done more confidently when Raman and XRD are used simultaneously. All calculated Raman spectra and the thermodynamic analysis code are published in open-access repositories alongside this work, allowing the adoption of our approach to other studies of $BaZrS_3$ synthesis.

## Methods:

### Experimental procedure:

Synthesis of powders: BaS (99.7%. Alfa Aesar) and $ZrS_2$ (99.99%, Alfa Chemistry) were used without further purification. 300 mg of powder formed by 1:1, 1:1.05, 1:1.1 and 1:1.2 molar mixtures of BaS and $ZrS_2$ (called Zr_0, Zr_5, Zr_10, Zr_20 respectively) were finely ground with an agate pestle and mortar, loaded into carbon-coated quartz ampules, purged with argon 3 times and sealed under a vacuum of $10^{-5}$ mbar. Each mixture was then placed in a single zone furnace at 500°C, and the temperature was increased to 900°C at a rate of 200°C/h. The powders were kept at 900°C for 5 days, and finally quenched in water.

### Computational details:

Competing phases of $BaZrS_3$ were identified using the Materials Project database,[23] with all Ba-Zr-S compounds within 0.5 eV above the convex hull considered. This energy range has been shown to cover the 90[th] percentile of all metastable materials reported within Materials Project.[24,25] First principles density functional theory (DFT) calculations were carried out with the all-electron numeric atom-centered code FHI-aims.[26] Minimum-energy crystal structures were found using parametrically constrained geometry relaxation.[27] The self-consistent field criteria was set to $10^{-7}$ e/Å$^3$ and $10^{-6}$ eV/Å for electron density and force respectively. The structures were relaxed until the maximum force component was below $5\times10^{-3}$ eV/Å. All other inputs were set to the default value within FHI-aims. All relaxations and phonon calculations were performed with the PBEsol[28] functional with a tight basis set. Electronic band structures and total energies were calculated using the HSE06[29] functional alongside inclusion of spin-orbit coupling. The resulting formation energies and band gaps are reported in Table S1. Phonon band structures were evaluated using the finite difference method with a 0.01 Å step size, as implemented in Phonopy.[30] The supercell size and k-point spacing used for phonon calculations are presented in Table S2. Raman intensities and peak positions were generated through Phonopy-Spectroscopy with a two-point finite difference along displacements, and are reported in the Supporting Information.[31] A Lorentzian peak width of 1 cm$^{-1}$ was set on the peak positions obtained from the analysis. Macroscopic dielectric tensors were evaluated using the real space density functional perturbation theory method implemented in FHI-aims[32] with the PBE functional.[33] Gibbs free energies were calculated using Phonopy and the ThermoPot package.[34] Imaginary modes were omitted from our calculation of the thermodynamic partition function and so do not contribute to the Gibbs free energy. An online repository containing i) analysis code for generating Figure 1a and Figure S2; ii) the raw data from electronic structure calculations; and iii) the data used to generate Raman spectra is available at https://github.com/NU-CEM/2022_BaZrS3_High-T_equilibrium.

**Results and discussion**

During BaZrS$_3$ synthesis competing phases can form from both the constituent elements (Ba, Zr, S) and/or from external impurities (e.g., O$_2$). In order to focus our analysis efforts on the thermodynamically accessible competing phases, ab-initio thermodynamic calculations were performed. As in this study synthesis was carried out within a closed ampule under vacuum, phases in the Ba-Zr-S system only were considered. Competing phases were initially identified using the Materials Project database,[35] followed by re-calculation of the total energies using a higher level of theory (HSE06 functional with spin-orbit coupling) to reproduce lattice constants and bandgaps in agreement with published results from experimental studies: the formation energy, lattice constants, electronic band gap and a comparison to experimental data for all systems considered are provided in the Table S1. Five competing ternary phases were identified: $Ba_4Zr_3S_{10}$ (*I4/mmm*), $Ba_3Zr_2S_7$ (*I4/mmm*), $Ba_3Zr_2S_7$ (*P42/mnm*), $Ba_3Zr_2S_7$ (*Cmmm*) and $Ba_2ZrS_4$ (*I4/mmm*). All ternary compounds are in the Ruddlesden Popper perovskite analogue series, $Ba_{n+1}Zr_nS_{3n+1}$.

It follows that there are three ternary-to-ternary decomposition mechanisms to consider:

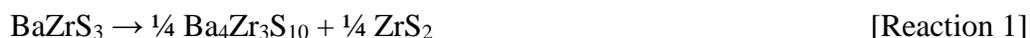

$BaZrS_3 \rightarrow ¼\ Ba_4Zr_3S_{10} + ¼\ ZrS_2$ [Reaction 1]

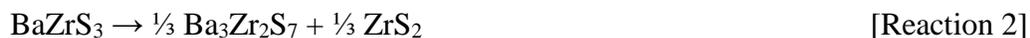

$BaZrS_3 \rightarrow ⅓\ Ba_3Zr_2S_7 + ⅓\ ZrS_2$ [Reaction 2]

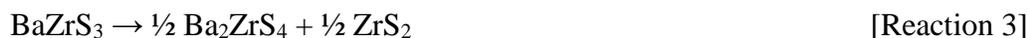

$BaZrS_3 \rightarrow ½\ Ba_2ZrS_4 + ½\ ZrS_2$ [Reaction 3]

To predict the relative phase stabilities a previously published methodology was followed to calculate the change in Gibbs free energy (ΔG) of each process,[36,37] as this is the potential that is minimised in equilibrium. It is important to emphasize that this calculation does not consider the effects of lattice expansion or anharmonic vibrations. However, despite these limitations, this methodology has been used to successfully predict the temperature-pressure stability window for $Cu_2ZnSnS_4$.[36]

The ΔG for all processes, reported in Figure 1a, is endothermic (positive valued) across the full temperature range considered (100K - 1300K). ΔG for [Reaction 3] ranges from 17kJ/mol to 15kJ/mol suggesting that $BaZrS_3$ is stable against decomposition into $Ba_2ZrS_4$. ΔG for [Reaction 2] is calculated for the three polymorphs of $Ba_3Zr_2S_7$. At 300K ΔG is comparable for all three polymorphs, as expected given the structural similarity: 10kJ/mol (*P42/mnm* and *Cmmm*), 11kJ/mol (*I4/mmm*). At 1300K ΔG is reduced to very small values: 7kJ/mol (*P42/mnm*), 6kJ/mol (*I4/mmm*) and 5kJ/mol (*Cmmm*). ΔG for [Reaction 1] is 10kJ/mol at RT reducing to 2kJ/mol at 1300K. Given that chemical accuracy is ~4kJ/mol, these results indicate that i) metastable $Ba_3Zr_2S_7$ and $Ba_4Zr_3S_{10}$ are energetically accessible during synthesis, with $Ba_4Zr_3S_{10}$ being the most likely phase to form ii) formation of $Ba_3Zr_2S_7$ and $Ba_4Zr_3S_{10}$ becomes more likely as the temperature is increased. Our results are supported by reports of $Ba_4Zr_3S_{10}$ formation during high-temperature synthesis of $Ba_3Zr_2S_7$.[6]

Crystal, electronic and vibrational structure information for $BaZrS_3$ and for the lowest energy competing phase, $Ba_4Zr_3S_{10}$, are shown in Figure 1b-g. The electronic band structure is calculated with the HSE06 functional and includes spin-orbit effects, leading to an accurate predicted value of 1.72eV[38] for the 3D perovskite direct band gap and a predicted value of 1.13 eV for the indirect band gap of the RP phase. This is the smallest predicted band gap for RP $Ba_{n+1}Zr_nS_{3n+1}$ materials considered, and is in line with previous reports of a decreasing band gap with increasing n.[20-22] Whilst an indirect bandgap can lead to a decreased absorption coefficient near the band edge, we expect this to be counter-balanced by the flatter band dispersion for $Ba_4Zr_3S_{10}$, resulting in a larger density of states.

At room temperature $BaZrS_3$ is reported to form in in space group *Pnma*,[9] which is a distortion of the idealised cubic perovskite. RP phases are reported to form in the space group *I4/mmm*.[20-21] It is found that $BaZrS_3$ in the *Pnma* phase is dynamically stable with positive phonon modes across the Brillouin zone. In contrast, $Ba_4Zr_3S_{10}$ is dynamically unstable with imaginary phonon modes at the zone boundaries. This indicates the presence of a symmetry lowering transition to a more stable phase at low-temperature, which is a common feature of halide and oxide perovskite materials.[39,40] $Ba_4Zr_3S_{10}$ in the space group *Fmmm* has also been previously reported in the literature.[18] However this corresponds to a small lattice expansion and increase in the c/a lattice parameter ratio, so is unlikely to result from distortions along zone boundary phonon modes of the *I4/mmm* phase. In Figures S1 we confirm that the high temperatures structures for $Ba_4Zr_3S_{10}$ (*Fmmm*), $Ba_3Zr_2S_7$ (*I4/mmm*) and $Ba_2ZrS_4$ (*I4/mmm*) also produce imaginary phonon modes in the harmonic approximation.

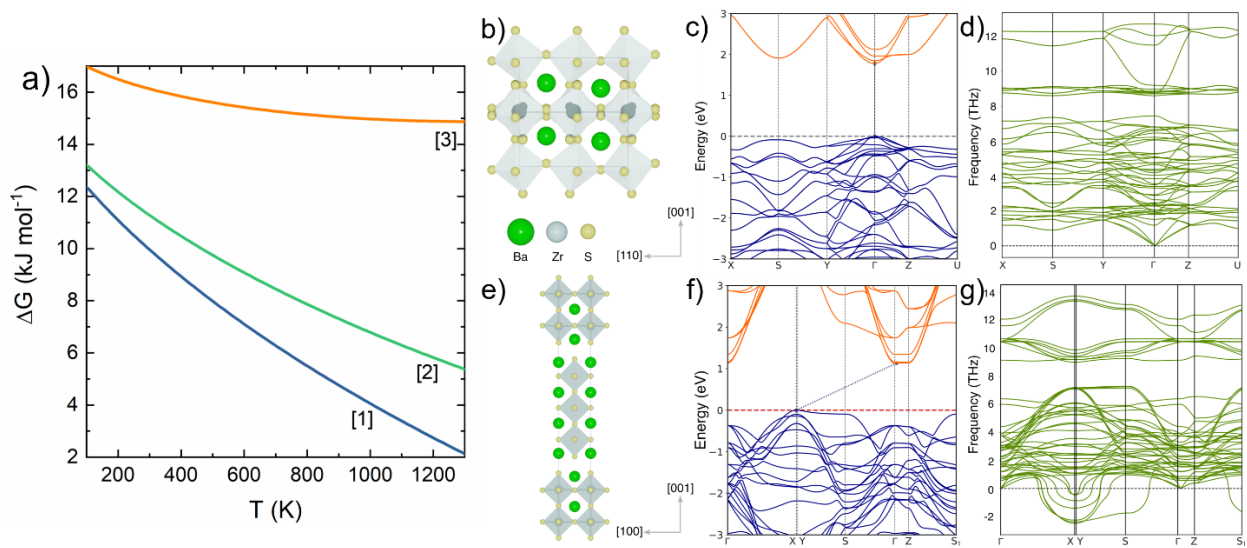

*Figure 1*: Thermodynamic, electronic and vibrational properties of BaZrS₃ (Pnma) and Ba$_{n+1}$Zr$_n$S$_{3n+1}$ (I4/mmm). a) Gibbs free energy ΔG as a function of Temperature. ΔG is calculated for decomposition of BaZrS₃ into: [1] Ba₄Zr₃S₁₀ and ZrS₂; [2] Ba₃Zr₂S₇ and ZrS₂; [3] Ba₂ZrS₄ and ZrS₂. For comparison, all materials are in the I4-mmm phase b) Crystal structure of BaZrS₃ c) Electronic band structure of BaZrS₃ calculated using the HSE06 exchange-correlation functional and with spin-orbit coupling (SOC), showing

*a direct band gap of 1.72eV[38] d) Phonon band structure of BaZrS$_3$ with positive phonon modes across the Brillouin Zone, indicating dynamical stability. e) Crystal structure of Ba$_4$Zr$_3$S$_{10}$ f) Electronic band structure of Ba$_4$Zr$_3$S$_{10}$ (HSE06+SOC), showing an indirect band gap of 1.13 eV g) Phonon band structure of Ba$_4$Zr$_3$S$_{10}$ showing negative phonon modes at the zone boundaries, and indicating the presence of a lower symmetry structure at 0T.*

To characterise the bulk structure, XRD was performed on powder samples at room temperature and ambient atmosphere. The corresponding diffractograms are reported in Figure 2a. The names Zr_0, Zr_5, Zr_10 and Zr_20 are used in this work represent the product obtained by the reaction of 1:1, 1:1.05, 1:1.1, 1:1.2 BaS:ZrS$_2$ molar ratios respectively.

Observing the diffractograms it can be noted that the samples present high crystallinity and, except the Zr_20, present complete conversion to the ternary phase, with no evidence of unreacted BaS (Figure S3a). The presence of unreacted ZrS$_2$ is more challenging to assess, as its characteristics peak is located at 32.2° (Figure S3a), where the samples present numerous small peaks which could derive also from the 2D or 3D perovskite. As discussed later in the text, Raman analysis simplified the assignation, confirming the presence of ZrS$_2$ in the Zr_0 sample. The Zr_20 sample presents an additional peak at low angles which origin has not been identified. Given the poor quality of the XRD pattern of this sample, both in terms of crystallinity and phase purity, it has been excluded from further characterization. It has to be noted that the presence of other Ba-S and Zr-S binary compositions have been excluded observing their XRD patterns and Raman spectra (Figure S3b). Similarly, the presence of oxides, sulfates and carbonates (all possible unwanted products in presence of air) have been excluded by comparing reported experimental XRD patterns and Raman spectra (Table S3).

The Zr_0 sample, which was expected to provide stoichiometric conversion to BaZrS$_3$,[41] shows the presence of multiple phases that were assigned to the 3D BaZrS$_3$ and to the RP phase Ba$_4$Zr$_3$S$_{10}$. It is important to highlight the similarity of the angular position and intensity of the BaZrS$_3$, Ba$_3$Zr$_2$S$_7$ and Ba$_4$Zr$_3$S$_{10}$ XRD patterns, which complicates the experimental diffractograms resolution (Figure S4). The Rietveld refinement presented in Figure S5 and Table S4-6 shows that the RP phase is the majority one, representing almost the 70% by mass. As the molar amount of ZrS$_2$ in the starting mixture is increased by 5%, the diffractogram still reveals a mixture of phases, but with an overall predominance of the BaZrS$_3$ over the RP phase.

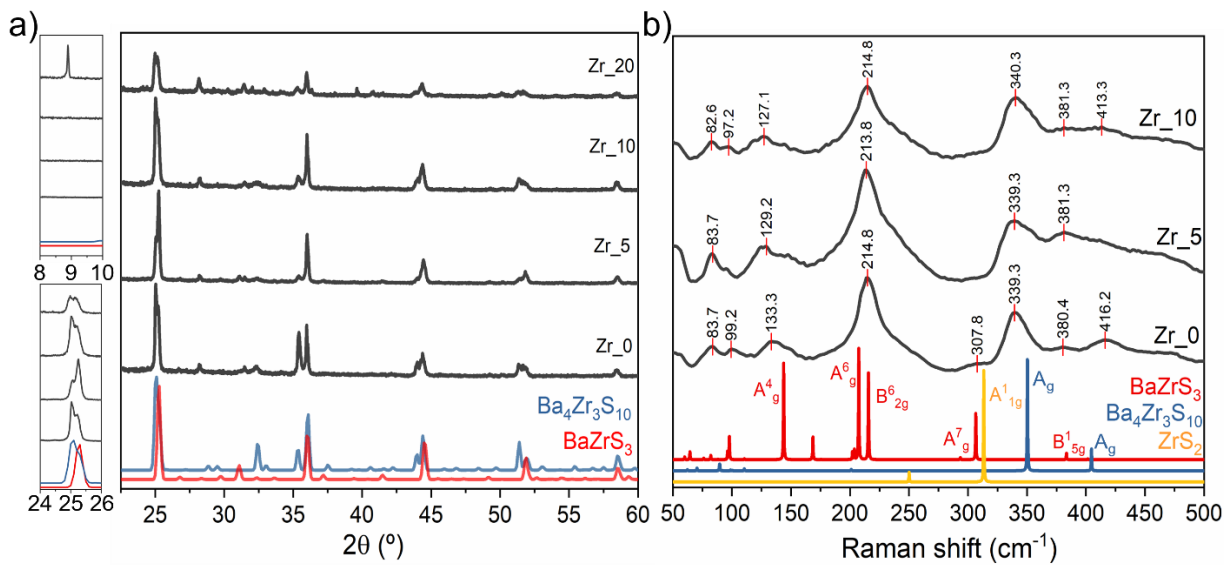

*Figure 2 a) X-Ray diffractograms of the synthesized powders. Reference patterns for BaZrS$_3$, Ba$_4$Zr$_3$S$_{10}$ are taken from ICSD (Collection Code 23288 and 72656 respectively). Top left panel: focus on the low angle peak shown by Zr_20. Bottom left panel: focus of the main peak. b) Calculated (bottom) and measured (top) Raman spectra collected with 785 nm excitation wavelength*

This can be concluded by the change in shape of the peak at 25.1° and by observing the high angle peaks (Figure S4), and is confirmed by the Rietveld refinement. For the sample with a nominal 10% molar excess of ZrS$_2$, Ba$_4$Zr$_3$Z$_{10}$ returns to be the majority phase, surprisingly reaching 86% of the sample weight.

*Table 1. Phase quantification from Rietveld refinement*

|       | BaZrS$_3$ (*Pnma*) %wt | Ba$_4$Zr$_3$S$_{10}$ (*Fmmm*) %wt | BaZrS$_3$ (*Pnma*) %mol | Ba$_4$Zr$_3$S$_{10}$ (*Fmmm*) %mol |
|-------|---|---|---|---|
| **Zr_0**  | 31 | 69 | 61 | 39 |
| **Zr_5**  | 59 | 41 | 84 | 16 |
| **Zr_10** | 14 | 86 | 36 | 64 |

In Figure 2b and Table 2 the experimentally recorded Raman spectra of all the samples are reported together with the calculated spectra of BaZrS$_3$, Ba$_4$Zr$_3$S$_{10}$ and ZrS$_2$. First of all, it is worth stressing that the Raman spectra allows clear distinction between the binary sulphides, the 2D RP phases and 3D CPVK, in contrast with the XRD pattern in Figure 2a. At a first look it can already be appreciated that in all the samples there is BaZrS$_3$, as shown by the most intense peak located at 214 cm$^{-1}$, deriving from the A$^6_g$ and B$^6_{2g}$ vibrational modes. Similarly, the peaks at lower angles (83 and 133 cm$^{-1}$) and the peak at 380 cm$^{-1}$ are mainly assigned to the BaZrS$_3$ phase ($A^2_g$, $A^4_g$ and $B^5_{1g}$ respectively), in agreement with previous works (Table 2).[5,40,41] It is worth stressing that the expected experimental spectra measured at room temperature should be red-shifted to lower frequencies compared to computational predictions, due to thermal expansion of the lattice which leads to a softening of the phonon modes. For the $A^3_g$ and $A^4_g$ modes in BaZrS$_3$ the frequency shift at room temperature is reported to be approximately 8 cm$^{-1}$,[42] in agreement with what observed (a shift of 10.3 cm$^{-1}$ and 6.6 cm$^{-1}$ respectively). In contrast, the experimental spectra are blue-shifted for the $A^6_g$ and $B^6_{2g}$ modes, as has been reported previously.[42,44] Importantly, these modes relate to displacements of the sulphur species only. In the sample Zr_0 a shoulder can be observed around 308 cm$^{-1}$, which is assigned to both the $A^7_g$ mode of the BaZrS$_3$ as well as the $A^1_{1g}$ of the ZrS$_2$. Since previously reported Raman spectra of BaZrS$_3$ show that the $A^7_g$ mode is visible only at low temperature,[42] the 308 cm$^{-1}$ peak in the Zr_0 spectrum is likely to derive from unreacted ZrS$_2$. This demonstrates that the presence of ZrS$_2$ is easier to observe through Raman spectroscopy than through XRD.

Importantly, all the spectra present a peak at around 339 cm$^{-1}$ which does not derive from the 3D perovskite. Comparing the measured spectrum with the calculated one for Ba$_4$Zr$_3$S$_{10}$ it is possible to assign it to the secondary RP phase, which is also corroborated by XRD. The Zr_0 sample presents an additional peak at 416 cm$^{-1}$, which can be assigned to the A$_g$ mode of the Ba$_4$Zr$_3$S$_{10}$. This phonon mode involves displacement of the sulphur species only, as in the $A^6_g$ and $B^6_{2g}$ modes of the BaZrS$_3$ phase, which can be used to rationalise the unexpected blue shift to higher frequencies for the experimental spectra.

*Table 2. Raman peak positions for the measured Zr_0 ($v_{exp}$) compared to the calculated value ($v_{pbesol}$).*

| Material | Mode | $v_{pbesol}$ | $v_{exp}$ |
|---|---|---|---|
| BaZrS$_3$ | $A^3_g$ | 94.0 | 83.7 |
| | $A^4_g$ | 139.8 | 133.3 |
| | $A^6_g$ | 204.9 | 214.8 |
| | $B^6_{2g}$ | 213.1 | 214.8 |
| | $B^5_{1g}$ | 389.1 | 380.4 |
| ZrS$_2$ | $A^1_{1g}$ | 313.4 | 307.8 |
| Ba$_4$Zr$_3$S$_{10}$ | A$_g$ | 350.1 | 339.3 |
| | A$_g$ | 403.9 | 416.2 |

Importantly, Raman spectra cannot be used for quantitative analysis unless calibration with phase pure materials is performed. The higher intensity of the BaZrS$_3$ Raman peaks compared to the RP ones does not indicate a higher content in the sample, as demonstrated by the Rietveld quantifications. However, it is interesting to note how the ratio between the areas of the 339 cm$^{-1}$ and 214 cm$^{-1}$ peak decreases as 5% molar excess of $ZrS_2$ is used in the starting mixture, suggesting that in this sample the orthorhombic 3-dimensional phase is favoured over the 2D. This hypothesis is supported by the disappearance of the peak at 99 cm$^{-1}$ (which does not derive from $BaZrS_3$ and is attributed to the $Ba_4Zr_3S_{10}$ mode located at 92 cm$^{-1}$) and to the increased intensity of the 380 cm$^{-1}$ peak. In the Zr_10 sample, instead, the ratio between the two peaks returns in favour to the $Ba_4Zr_3S_{10}$, with a reduction of the $BaZrS_3$ peaks, as confirmed by Rietveld analysis. It should be noted that the measured Raman spectra were compared against the spectra of the binary precursors (Figure S3) as well as against the oxide counterparts (Table S3),[45–49] confirming their absence in the synthesised mixture.

So far it has been shown that, in these synthetic conditions, the reaction between binary precursors leads to the formation of a secondary 2D ternary phase. Zr- or S-poor conditions can trigger the formation of RP phases, and a careful compositional analysis is necessary to obtain sensible conclusions. For this reason EDS compositional analysis has been performed on all the synthesized powders and on the Zr_0 precursor mixture before heat treatment, and the results are reported in Table 3.

*Table 3: EDS quantifications (in atomic %) of the Zr_0 powder before the synthesis and after, Zr_5 and Zr_10 with the instrumental error.*

|                       | Ba      | Zr      | S       |
|-----------------------|---------|---------|---------|
| **Zr_0 before treatment** | 23 ± 5  | 22 ± 5  | 55 ± 5  |
| **Zr_0 after treatment**  | 25 ± 5  | 21 ± 5  | 54 ± 5  |
| **Zr_5**              | 25 ± 5  | 22 ± 5  | 53 ± 5  |
| **Zr_10**             | 25 ± 5  | 23 ± 5  | 52 ± 5  |

The Zr_0 precursor mixture confirms that the Ba/Zr atomic ratio before heat treatment was 1:1. Surprisingly, all the resulting powders after treatment present Zr deficient compositions, with the Ba/Zr atomic ratio decreasing in the Zr_0, Zr_5 and Zr_10 series. However it is worth stressing that the confidence range of these values is enlarged by the limitations associated with powders analysis with EDS (see Supporting Information). The starting material and the synthesised powders also present sub-stoichiometric amounts of sulphur, which become more unbalanced as the starting excess of $ZrS_2$ is higher. To exclude accidental losses during the loading of the samples in the quartz ampules, a second batch repeating the Zr_0 conditions was prepared and characterized showing very similar XRD, Raman spectra and EDS compositions to Zr_0 (Figure S6). The systematic loss of Zr in these synthesis hints to possible unwanted reactions with the quartz ampule, even if carbon-coated. More investigation is needed to address this phenomenon, which is out of the scope of this work. However this evidence hints that other precursors, rather than $ZrS_2$, should be used, especially in high-temperature solid-state synthesis.

The observation of sub-stoichiometric amounts of Zr and S gives an additional explanation for the formation of $Ba_4Zr_3S_{10}$. As the RP $Ba_{n+1}Zr_nS_{3n+1}$ family of materials are a $ZrS_2$-deficient analogue of $BaZrS_3$, the RP phases are expected to be more readily formed in Zr- or S-poor environments. To explore this further, a ternary phase diagram for the Ba-Zr-S system was constructed using the first-principles thermodynamic model introduced earlier (Figure S2). This allows a prediction of which products are formed for the sub-stoichiometric amounts of Zr and S as measured by EDS. At high temperature all RP phases and most binary phases (all except $BaS_3$) lie on the convex hull, so that the predicted products are very sensitive to composition. Using Ba-Zr-S composition values within the range of values reported for each sample in Table 3, the model predicts that $Ba_4Zr_3S_{10}$ and $BaZrS_3$ will be formed at at 900°C, alongside a smaller proportion of $ZrS_2$ (Table S7). This suggests that there is an additional driving force for phase separation into $BaZrS_3$ and $Ba_4Zr_3S_{10}$ resulting from the under stoichiometric amount of Zr and S. Our modelling also suggests that the S concentration in the samples limits the formation of $BaZrS_3$. This is why the Zr_10 sample, with the lowest proportion of sulphur measured by EDS, produces the highest proportion of $Ba_4Zr_3S_{10}$. On the other hand, this sample has the Ba/Zr ratio closer to unity, suggesting that the cation ratio is not the driving force for the preferential formation of 3D over 2D perovskites.

The compositional mapping presented in Figure 3 offers an additional insight in the nature of the synthesised powders. In the image, brighter zones correspond to areas of high atomic concentration, although it is recognised that shadowing effects give large dark areas (observable in the dark large areas in the mappings but not in the secondary electron images). Notwithstanding this, it may be observed that in all the samples Zr is less uniformly distributed than Ba and S, as evidenced by the particularly bright area in the Zr mapping. Among all the samples, the Zr_0 shows the biggest agglomeration of zirconium. Brighter area can be observed also in the other samples, but the dimension and number of these agglomerates reduces in Zr_5 and Zr_10 with the formation of smaller clusters more dispersed across the sample. The EDS analysis confirmed the presence of Zr-rich clusters in the sample Zr_0 (Figure S7), but in the others the distribution of Zr looks more homogeneous (Figure S8 and S9), even if some local deviations from the average are still present, as indicated by the mapping results. Instead, Ba and S mapping show that the Zr-rich areas in the Zr_0 sample correspond to S and Ba deficient zones. The presence of Zr-rich clusters hinders the uniformity of the reaction environment possibly creating local regions even more Zr-deficient than the average composition, favouring the formation of 2D perovskites. Furthermore, the morphology of the synthesised crystals has been analysed through scanning electron microscopy (SEM) in both secondary electron and the more compositionally sensitive backscattered electron (BSE) modes. Looking at the BSE image in Figure S10 it can be noted that, in the Zr_0 sample, the crystals have a different appearance accordingly with the grain composition, as it could also be appreciated in Figure S7. The grains with a composition close to stoichiometry have a flat, uniform surface, while the grains presenting high concentrations of Zr present an irregular, rough, patchy microstructure within the grain. Additionally, it can be observed that the flat and regular grains present a laminar structure, further indication of the presence of layered 2D phases in the sample.

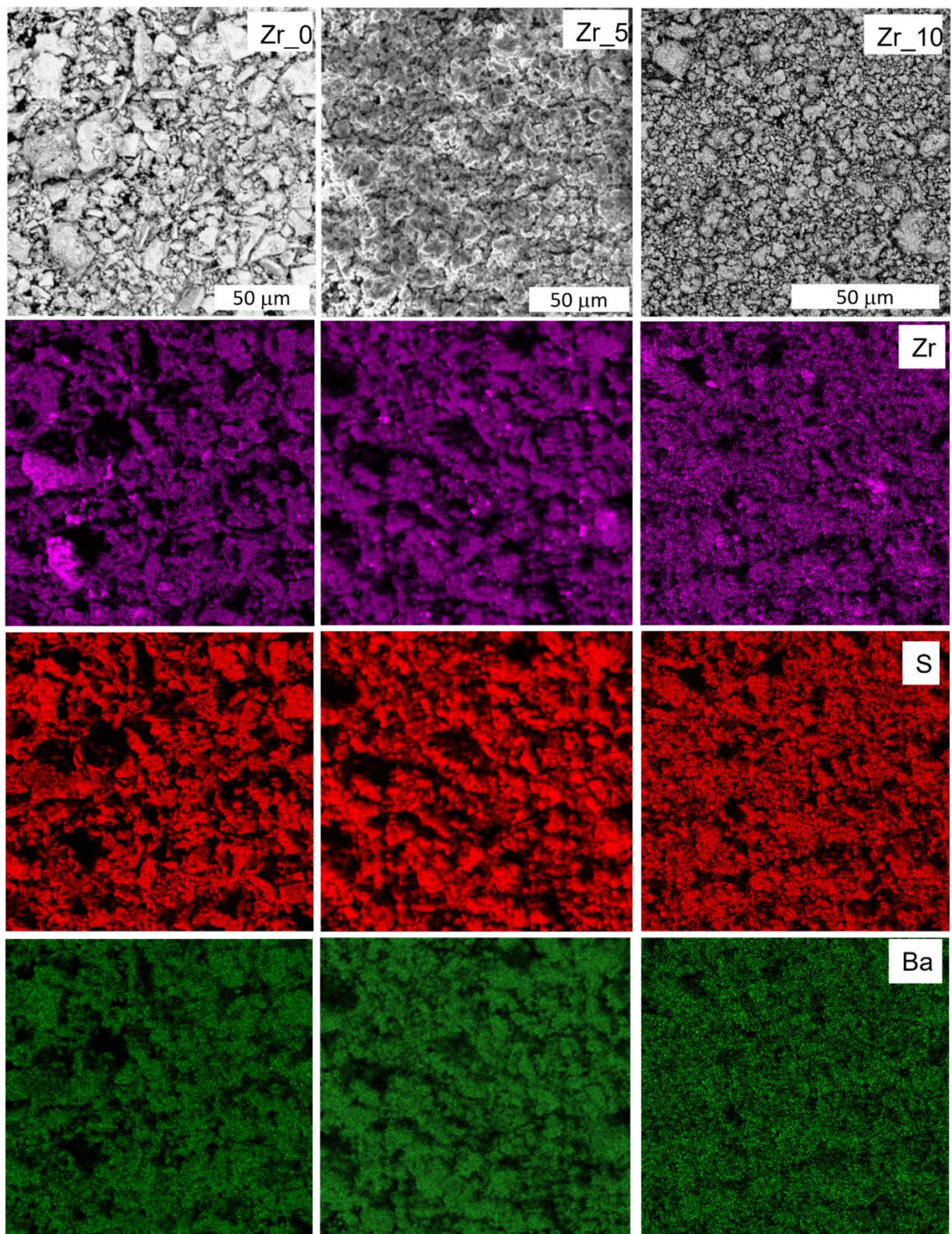

*Figure 3. Secondary electron images and elemental mapping of the Zr_0, Zr_5 and Zr_10 (left, centre and right column respectively)*

**Conclusions**

In this work experimental and computational techniques have been combined to identify structurally-similar products of the reaction between BaS and $ZrS_2$ at different stoichiometric ratios. Ab-initio thermodynamic calculations identified that chalcogenide RP phases (and in particular the phases with larger n) are energetically close to $BaZrS_3$ and are thermodynamically accessible at the high temperatures often required for the CPVK synthesis. XRD and Raman spectroscopy demonstrate that for all the investigated ratios of binary chalcogenide precursors, mixtures of 3D $BaZrS_3$ and 2D $Ba_4Zr_3S_{10}$ are created. Interestingly it has been shown that, despite the use of $ZrS_2$ excess in the precursor mixtures, Zr and S deficiencies have been observed in all the synthesised powders, further driving the 2D phase formation. However, the formation of RP phase does not depend only on the Ba/Zr ratio (with higher concentration of 2D phases formed with the Ba/Zr ratio closer to unity), but is also extremely sensitive to sulphur deficiencies. These results suggest the high possibility of creation of mixed phases during CPVK synthesis at high temperatures, and the need to carefully control phase composition. These results may also explain why the reported band gap of $BaZrS_3$ varies significantly in different publications: it is possible that 3D and 2D mixtures were created and not identified due to the similarity of the XRD diffractograms of these species.

Raman spectra calculated from first principles have been published in an open-source database for binary Ba-S and Zr-S and ternary Ba-Zr-S compositions. This includes Raman data not yet present in the literature, specifically for phases $Ba_4Zr_3S_{10}$, $BaS_2$, $BaS_3$, ZrS and $ZrS_3$. $Ba_4Zr_3S_{10}$ has been rarely reported, and little characterization is available in literature on this RP perovskite. The DFT calculations presented predict an indirect bandgap of 1.13 eV, suggesting that it the smallest band gap of materials synthesised in the $Ba_{n+1}Zr_nS_{3n+1}$ series.

Finally, this work has demonstrated the complexity of the Ba-Zr-S phase diagram, stressing the importance of using multi-experimental techniques to soundly resolve the reaction products of synthesis, and the necessity to find alternative synthetic routes involving lower temperatures and different precursors. Raman spectroscopy, here suggested as a technique complimentary to XRD, has been proven to provide a relatively easy and quick differentiator between CPVK phases, and even for several binary compositions of the Ba-Zr-S elements. In addition, the use of Raman spectroscopy in other CPVK synthesis environments (such as nanoparticle synthesis) may help to identify the reaction mechanisms where traditional techniques, such as XRD, cannot be used, or where the presence of organic materials complicates the analysis. Hopefully, the use of the database created in this work will support further progress in developing a low-temperature synthetic procedure for chalcogenide perovskites, promote increased control of the 2D-3D phase equilibrium during synthesis, and ultimately enable their thin film deposition and integration into optoelectronic devices.

**Acknowledgements**

PK and LDW thank Jonathan M Skelton for discussions on Phonopy-Spectroscopy. PK and GL acknowledges support from the UK Engineering and Physical Sciences Research Council (EPSRC) CDT in Renewable Energy Northeast Universities (ReNU) for funding through EPSRC Grant EP/S023836/1. The authors thanks Horiba for the support with the Raman measurements. This work used the Oswald High Performance Computing facility operated by Northumbria University (UK). Via our membership of the UK's HEC Materials Chemistry Consortium, which is funded by EPSRC (EP/R029431), this work used the ARCHER2 UK National Supercomputing Service (http://archer2.ac.uk).

**Data availability**

Raman plots, peak tables and symmetry analysis of all mentioned phases are provided in the supplementary material. All raw data needed to reproduce the results have been uploaded to Zenodo.

**Conflict of interest**

The authors declare no conflict of interest.

Prakriti Kayastha, Devendra Tiwari, Adam Holland, Oliver S. Hutter, Ken Durose, Lucy D Whalley* and Giulia Longo*

**High temperature equilibrium of 3D and 2D chalcogenide perovskites.**

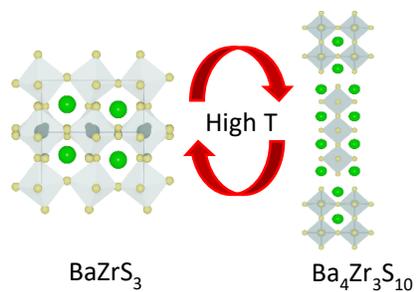

BaZrS$_3$    Ba$_4$Zr$_3$S$_{10}$

At high temperatures the formation energy of 3D and 2D chalcogenide perovskites are very close. Mixture of phases is then formed in these conditions.

# Supporting information

**High temperature equilibrium of 3D and 2D chalcogenide perovskites.**

*Prakriti Kayastha, Devendra Tiwari, Adam Holland, Oliver S. Hutter, Ken Durose, Lucy D. Whalley\* and Giulia Longo\**

**X-Ray Diffraction**

The powder X-ray diffraction patterns were acquired using Rigaku SmartLab SE, in Bragg Brentano geometry and with Cu Kα1 (1.54056 Å) radiation. The diffractograms were collected from 7 to 80°, with long acquisition time (0.2°/min) to have sufficient angular resolution for Rietveld refinement.

**Rietveld Analysis**

Full-profile refinement and quantitative phase analysis was performed using Rietveld method as implemented in FullProf suite.[1,2]

**Raman Spectroscopy**

Raman spectra were collected using Horiba LabRAM Soleil system, equipped with a 785 nm laser. The power at the sample was measured to be 4.7 mW. The used objective was a 50X with a 11mm working distance, the confocal hole was 200 mm and the grating 1200 g/mm.

**Scanning Electron Microscopy and Energy Dispersive X-Ray (SEM-EDS)**

Scanning electron microscopy images were acquired using Tescan Mira 3 FEG-SEM, employing both secondary electron and back-scattered electron detectors. Energy dispersive Xray and elemental mapping were collected using a X-Max detector and Aztec software (Oxford Instruments), using 20KV electron acceleration and ZAF correction algorithm. The elemental compositions presented in Table 3 were obtained by averaging the values obtained over 10 different areas (600*600 µm) for each sample. These values are reported with a 5% uncertainty due to the nature of the measurement: EDS can be a very accurate technique to measure sample composition, provided that the sample has a flat and smooth surface, uniform thickness and density, and that a calibration standard for each element of interest is measured together with the samples. The powders analysed in this work presented very rough surfaces and strong variation in the sample thickness. The morphological characteristics of the samples can strongly affect the accuracy of the measurements, due to electron scattering and X-ray re-absorption. Additionally, analysis was done without standards for Ba, Zr, S (even if quantification were been verified on other standard samples). For this reason, even if precise measurements were carried out (ie, low standard deviation between the 10 points analysed for each sample, ranging around 0.5%), the overall accuracy is reduced by the standardless analysis, and for this reason it was assigned to 5%.

WILEY-VCH**First-principles predictions of Raman peak position and intensity**

Calculation details are included in the main text.

**ZrS$_2$ (P-3m1)**

| v(cm$^{-1}$) | I (Å$^4$amu$^{-1}$) | Symmetry |
|---|---|---|
| 250.129723 | 97.526512 | E$^1$g |
| 250.129723 | 97.266348 | E$^1$g |
| 313.394668 | 2191.392129 | A$^1{}_{1g}$ |

**ZrS$_3$ (P2$_1$/m)**

| v(cm$^{-1}$) | I (Å$^4$amu$^{-1}$) | Symmetry |
|---|---|---|
| 73.014879 | 6.905896 | A$^1$g |
| 92.527391 | 1.166059 | B$^1$g |
| 110.387709 | 4.719362 | A$^2$g |
| 140.243367 | 250.082279 | A$^3$g |
| 141.667912 | 87.993663 | B$^2$g |
| 217.777387 | 311.319123 | B$^3$g |
| 240.498025 | 239.793097 | B$^4$g |
| 270.562547 | 46.823381 | A$^4$g |
| 273.792884 | 981.725127 | A$^5$g |
| 313.414725 | 1356.801371 | A$^6$g |
| 324.054591 | 183.217121 | A$^7$g |
| 514.353526 | 10802.39804 | A$^8$g |

**ZrS (P4/nmm)**

| v(cm$^{-1}$) | I (Å$^4$amu$^{-1}$) | Symmetry |
|---|---|---|
| 35.001975 | 282918.7229 | E$^1$g |
| 35.001975 | 282918.674 | E$^1$g |
| 185.135088 | 12149.69593 | A$^1{}_{1g}$ |
| 258.307259 | 523466.9858 | A$^2{}_{1g}$ |
| 301.317526 | 34598.75679 | E$^2{}_g$ |
| 301.317526 | 34598.78312 | E$^2{}_g$ |

**ZrS (Fm-3m)**

Not Raman active

**$Zr_3S_4$ (Fd-3m)**

| v(cm$^{-1}$) | I (Å$^4$amu$^{-1}$) | Symmetry |
|---|---|---|
| 111.859677 | 18277.38917 | $T^1_{2g}$ |
| 111.859677 | 48563.4653 | $T^1_{2g}$ |
| 111.859677 | 16221.79423 | $T^1_{2g}$ |
| 209.735626 | 11.036357 | $T^2_{1g}$ |
| 209.735626 | 10.938889 | $T^2_{1g}$ |
| 209.735626 | 0.022322 | $T^2_{1g}$ |
| 233.997049 | 228529.875 | $E^1_g$ |
| 233.997049 | 12112.62315 | $E^1_g$ |
| 242.657169 | 5155.011086 | $T^3_{2g}$ |
| 242.657169 | 6326.453296 | $T^3_{2g}$ |
| 242.657169 | 2002.857961 | $T^3_{2g}$ |
| 384.674094 | 868.761597 | $T^4_{2g}$ |
| 384.674094 | 654.553594 | $T^4_{2g}$ |
| 384.674094 | 621.6211 | $T^4_{2g}$ |
| 396.328116 | 55408.12694 | $A^1_{1g}$ |

**BaS (Fm-3m)**

Not Raman active

**$BaS_2$ (C2/c)**

| v(cm$^{-1}$) | I (Å$^4$amu$^{-1}$) | Symmetry |
|---|---|---|
| 68.229845 | 2.857842 | $B^1_g$ |
| 82.348163 | 3.31171 | $A^1_g$ |
| 134.829312 | 2.210724 | $B^2_g$ |
| 164.413337 | 115.845998 | $A^2_g$ |
| 165.237729 | 67.569019 | $B^3_g$ |
| 178.574775 | 116.502308 | $B^4_g$ |
| 188.72554 | 716.275971 | $A^3_g$ |
| 446.184748 | 5377.260127 | $B^5_g$ |
| 448.277011 | 11368.46861 | $A^4_g$ |

**BaS$_3$ (P-42$_1$m)**

| v(cm$^{-1}$) | I (Å$^4$amu$^{-1}$) | Symmetry |
|---|---|---|
| 79.228808 | 1.055844 | B$_1$ |
| 81.827263 | 5.553188 | E |
| 81.827263 | 5.553188 | E |
| 98.185375 | 0.351295 | E |
| 98.185375 | 0.351295 | E |
| 103.481236 | 79.06596 | A$_1$ |
| 124.753145 | 3.761728 | B$_2$ |
| 154.037718 | 23.005822 | E |
| 154.037718 | 23.005822 | E |
| 164.530524 | 0.483868 | E |
| 164.530524 | 0.483868 | E |
| 194.410123 | 120.118113 | B$_1$ |
| 206.577304 | 3.097366 | E |
| 206.577304 | 3.097366 | E |
| 210.592472 | 0.000012 | A$_2$ |
| 219.278783 | 213.329742 | B$_2$ |
| 219.42517 | 491.262822 | A$_1$ |
| 438.629548 | 1319.955708 | B$_2$ |
| 444.395021 | 2483.759672 | A$_1$ |
| 458.725299 | 124.300156 | E |
| 458.725299 | 124.296929 | E |

**BaZrS$_3$ (Pnma)**

| v(cm$^{-1}$) | I (Å$^4$amu$^{-1}$) | Symmetry |
|---|---|---|
| 57.38246 | 31.294167 | A$^1_g$ |
| 62.740697 | 74.397711 | B$^1_{2g}$ |
| 72.865691 | 20.712113 | A$^2_g$ |
| 74.878717 | 0.428877 | B$^1_{3g}$ |
| 79.451252 | 35.213175 | B$^2_{2g}$ |
| 82.730798 | 4.047662 | B$^1_{1g}$ |
| 89.396457 | 81.906317 | B$^3_{2g}$ |
| 94.004782 | 202.648574 | A$^3_g$ |
| 105.117071 | 8.151943 | B$^2_{1g}$ |
| 139.776095 | 927.058612 | A$^4_g$ |

| v(cm⁻¹) | I (Å⁴amu⁻¹) | Symmetry |
|---|---|---|
| 157.205497 | 0.129784 | $B^2_{3g}$ |
| 162.91409 | 192.607928 | $B^4_{2g}$ |
| 166.327512 | 3.336429 | $A^5_g$ |
| 171.34857 | 2.338391 | $B^3_{1g}$ |
| 198.430549 | 98.18659 | $B^5_{2g}$ |
| 200.803735 | 93.792521 | $B^3_{3g}$ |
| 204.900935 | 1063.364972 | $A^6_g$ |
| 213.109461 | 766.788035 | $B^6_{2g}$ |
| 289.681849 | 19.314493 | $B^4_{1g}$ |
| 289.943548 | 8.201392 | $B^4_{3g}$ |
| 304.087743 | 361.586611 | $A^7_g$ |
| 389.08038 | 48.066782 | $B^5_{1g}$ |
| 400.954212 | 7.889141 | $B^7_{2g}$ |
| 419.617256 | 0.634279 | $B^5_{3g}$ |

**$Ba_2ZrS_4$ (I4/mmm)**

| v(cm⁻¹) | I (Å⁴amu⁻¹) | Symmetry |
|---|---|---|
| 52.406628 | 3.458753 | $E^1_g$ |
| 52.40262 | 3.459939 | $E^1_g$ |
| 114.218542 | 3.162731 | $A^1_{1g}$ |
| 121.15813 | 0.087169 | $E^2_g$ |
| 121.22848 | 0.08895 | $E^2_g$ |
| 314.503259 | 895.624663 | $A^2_{1g}$ |

**$Ba_3Zr_2S_7$ (P4$_2$/mnm)**

| v(cm⁻¹) | I (Å⁴amu⁻¹) | Symmetry |
|---|---|---|
| 30.212846 | 2.697724 | $E^1_g$ |
| 30.212846 | 2.697724 | $E^1_g$ |
| 38.927635 | 17.093892 | $E^2_g$ |
| 38.927635 | 17.093892 | $E^2_g$ |
| 50.204777 | 5.139854 | $E^3_g$ |
| 50.204777 | 5.139854 | $E^3_g$ |
| 50.316629 | 3.236544 | $B^1_{1g}$ |
| 52.517914 | 0.000001 | $A^1_{2g}$ |
| 61.681258 | 29.34648 | $A^1_{1g}$ |
| 62.882698 | 34.818208 | $B^1_{2g}$ |
| 64.636587 | 0.088883 | $B^2_{2g}$ |
| 68.110901 | 9.90813 | $B^2_{1g}$ |
| 68.438659 | 2.357514 | $E^4_g$ |
| 68.438659 | 2.357514 | $E^4_g$ |
| 68.558261 | 0.000011 | $A^2_{2g}$ |
| 71.395331 | 8.355441 | $B^3_{1g}$ |
| 71.439568 | 11.715276 | $E^5_g$ |
| 71.439568 | 11.715276 | $E^5_g$ |
| 73.772491 | 81.142077 | $A^2_{1g}$ |
| 79.402247 | 3.381258 | $E^6_g$ |
| 79.402247 | 3.381258 | $E^6_g$ |
| 80.183165 | 5.188376 | $B^3_{2g}$ |
| 85.101877 | 0.000096 | $A^3_{2g}$ |
| 86.888262 | 10.549098 | $E^7_g$ |
| 86.888262 | 10.549098 | $E^7_g$ |
| 91.231012 | 143.350971 | $A^3_{1g}$ |
| 103.480549 | 640.233835 | $A^4_{1g}$ |
| 106.849569 | 71.574558 | $E^8_g$ |
| 106.849569 | 71.574558 | $E^8_g$ |
| 110.939157 | 151.224356 | $B^4_{2g}$ |
| 112.446944 | 10.2548 | $E^9_g$ |
| 112.446944 | 10.2548 | $E^9_g$ |
| 119.000374 | 221.437353 | $A^5_{1g}$ |
| 127.095916 | 12.141797 | $B^5_{2g}$ |
| 127.641234 | 0.478178 | $E^{10}_g$ |
| 127.641234 | 0.478178 | $E^{10}_g$ |
| 135.11567 | 25.034852 | $E^{11}_g$ |
| 135.11567 | 25.034852 | $E^{11}_g$ |
| 136.820878 | 65.920617 | $B^4_{1g}$ |
| 140.631532 | 38.959964 | $A^6_{1g}$ |
| 140.891116 | 66.708777 | $E^1_{2g}$ |
| 140.891116 | 66.708777 | $E^1_{2g}$ |
| 146.007681 | 0 | $A^4_{2g}$ |
| 146.279277 | 3.635712 | $B^6_{2g}$ |
| 156.046509 | 139.804471 | $A^7_{1g}$ |

| v(cm⁻¹) | I (Å⁴amu⁻¹) | Symmetry |
|---|---|---|
| 159.535313 | 2.659249 | $B^7_{2g}$ |
| 190.33959 | 563.31957 | $A^8_{1g}$ |
| 191.77194 | 552.223312 | $B^8_{2g}$ |
| 199.204596 | 1.253092 | $B^5_{1g}$ |
| 199.741278 | 0.000034 | $A^5_{2g}$ |
| 209.975536 | 18.081153 | $E^{13}_g$ |
| 209.975536 | 18.081153 | $E^{13}_g$ |
| 211.827231 | 1.408878 | $B^9_{2g}$ |
| 211.988927 | 246.081075 | $A^9_{1g}$ |
| 229.9484 | 59.019295 | $A^{10}_{1g}$ |
| 230.161149 | 5.776477 | $B^{10}_{2g}$ |
| 240.23487 | 13.410356 | $B^6_{1g}$ |
| 240.397813 | 0.000001 | $A^6_{2g}$ |
| 286.462652 | 3.595897 | $E^{14}_g$ |
| 286.462652 | 3.595897 | $E^{14}_g$ |
| 294.502698 | 63.08334 | $E^{15}_g$ |
| 294.502698 | 63.08334 | $E^{15}_g$ |
| 302.43342 | 6.091773 | $E^{16}_g$ |
| 302.43342 | 6.091773 | $E^{16}_g$ |
| 321.579069 | 10.229922 | $E^{17}_g$ |
| 321.579069 | 10.229922 | $E^{17}_g$ |
| 338.460731 | 9384.451005 | $A^{11}_{1g}$ |
| 341.510689 | 1.836837 | $E^{18}_g$ |
| 341.510689 | 1.836837 | $E^{18}_g$ |
| 359.672158 | 0.00138 | $B^{11}_{2g}$ |
| 434.663829 | 2.302284 | $E^{19}_g$ |
| 434.663829 | 2.302284 | $E^{19}_g$ |

**$Ba_3Zr_2S_7$ (I4/mmm)**

| v(cm⁻¹) | I (Å⁴amu⁻¹) | Symmetry |
|---|---|---|
| 28.596154 | 4.245601 | $E^1_g$ |
| 28.600797 | 3.724356 | $E^1_g$ |
| 65.941815 | 17.660479 | $E^2_g$ |
| 65.953727 | 19.651663 | $E^2_g$ |
| 93.516505 | 76.443641 | $E^3_g$ |
| 93.519081 | 76.982938 | $E^3_g$ |
| 96.890024 | 47.707665 | $A^1_{1g}$ |
| 105.537887 | 0.451137 | $E^4_g$ |
| 105.567227 | 1.207343 | $E^4_g$ |
| 129.915843 | 16.255944 | $A^2_{1g}$ |
| 135.676242 | 2.664061 | $B^1_{1g}$ |
| 207.422076 | 54.634636 | $A^3_{1g}$ |
| 344.859098 | 3083.281673 | $A^4_{1g}$ |
| 358.176047 | 0.303150 | $E^5_g$ |
| 358.177755 | 0.380385 | $E^5_g$ |

**$Ba_3Zr_2S_7$ (Cmmm)**

| $v(cm^{-1})$ | $I\ (Å^4 amu^{-1})$ | Symmetry |
|---|---|---|
| -19.238772 | 0.000045 | |
| 10.580514 | 0.000034 | |
| 23.722276 | 8.115649 | |
| 28.358265 | 0.001798 | |
| 28.659876 | 8.079071 | |
| 66.060054 | 36.454634 | $B_{1g}$ |
| 70.453963 | 53.852095 | $B_{1g}$ |
| 80.524067 | 0.00001 | |
| 86.17592 | 0.000007 | $A_g$ |
| 92.944235 | 153.091092 | $A_g$ |
| 93.110809 | 134.476747 | $B_{1g}$ |
| 96.729933 | 94.646368 | |
| 102.537768 | 0.000021 | $A_g$ |
| 105.453077 | 3.054152 | |
| 106.889221 | 4.282398 | $B_{2g}$ |
| 129.799904 | 32.426702 | $A_g$ |
| 134.393598 | 4.806921 | $B_{1g}$ |
| 165.055096 | 0.000061 | $B_{1g}$ |
| 167.37705 | 0.000025 | $B_{1g}$ |
| 170.236148 | 0.000062 | $A_g$ |
| 206.92542 | 0.000002 | $A_g$ |
| 207.257653 | 109.413592 | |
| 241.027107 | 0.000003 | $B_{1g}$ |
| 241.915904 | 0.000058 | |
| 301.291395 | 0.000002 | $A_g$ |
| 335.415703 | 0.00000 | |
| 344.597264 | 6167.55051 | $A_g$ |
| 358.051261 | 0.548091 | $B_{3g}$ |
| 358.247058 | 0.674951 | $B_{1g}$ |
| 450.974406 | 0.000006 | $B_{1g}$ |

**$Ba_4Zr_3S_{10}$ (I4/mmm)**

| $v(cm^{-1})$ | I ($Å^4amu^{-1}$) | Symmetry |
|---|---|---|
| 23.888283 | 6.202089 | |
| 24.136284 | 8.566226 | |
| 28.794929 | 0.422725 | |
| 28.913987 | 0.085075 | |
| 62.572685 | 23.143253 | |
| 63.943724 | 15.596888 | |
| 70.26505 | 97.731324 | $A_{1g}$ |
| 90.574107 | 156.564256 | |
| 92.529283 | 167.047616 | |
| 98.698999 | 31.244085 | |
| 101.757759 | 7.552974 | |
| 101.921106 | 8.699022 | |
| 110.352799 | 76.837932 | $A_{1g}$ |
| 125.791600 | 0.204991 | |
| 125.922965 | 0.085178 | |
| 126.140783 | 0.137591 | |
| 200.968021 | 57.552119 | $A_{1g}$ |
| 350.057467 | 4187.23342 | $A_{1g}$ |
| 350.789978 | 0.449881 | |
| 350.914053 | 0 | |
| 351.499144 | 7.675391 | |
| 403.926303 | 803.45614 | $A_{1g}$ |

**$Ba_4Zr_3S_{10}$ (Fmmm)**

| $v(cm^{-1})$ | I ($Å^4amu^{-1}$) | Symmetry |
|---|---|---|
| 23.320502 | 7.245854 | $B_{3g}$ |
| 23.539796 | 7.12991 | $B_{2g}$ |
| 28.634693 | 0.358348 | $B_{3g}$ |
| 28.780656 | 0.459457 | $B_{2g}$ |
| 62.221252 | 20.906022 | $B_{3g}$ |
| 62.323541 | 22.165325 | $B_{2g}$ |
| 70.376478 | 98.903215 | $A_g$ |
| 89.466405 | 163.720452 | $B_{2g}$ |
| 89.746283 | 162.866758 | $B_{3g}$ |
| 98.766874 | 31.336473 | |
| 100.446552 | 8.561731 | |
| 100.670285 | 7.43394 | |
| 110.474511 | 75.254317 | $A_g$ |
| 126.317534 | 0.28124 | |
| 127.205923 | 0.003466 | |
| 127.337609 | 0.000294 | $B_{3g}$ |
| 201.055524 | 57.317053 | $A_g$ |
| 350.345923 | 4221.70852 | |
| 350.800913 | 0.451014 | $B_{2g}$ |
| 350.896492 | 0.597932 | |
| 404.581634 | 805.655363 | $A_g$ |

*Table S1: Predicted properties for materials in the Ba-Zr-S system. Competing phases of $BaZrS_3$ were identified using the Materials Project database, with all Ba-Zr-S compounds within 0.5 eV above the convex hull considered. This was followed by a re-calculation of the total energies using the HSE06 functional with spin-orbit coupling. The energy above the hull, lattice constants and electronic band gap for all systems considered are provided, alongside comparison to experiment.*

| Material (Space Group) | Formation energy (this work) (eV/atom) | Lattice constants (Å) | | Calculated (this work) electronic band gap (eV) |
|---|---|---|---|---|
| | | Calculated (this work) | Experiment | |
| BaS (Fm-3m) | -2.374 | a = 6.32, | a = 6.387[3] | 2.956 (indirect) |
| $BaS_2$ (C2/c) | -1.636 | a = 4.832, b = 9.3518, c = 9.5004, | a = 4.736, b = 8.993, c = 9.299[4] | 2.557 (indirect) |
| $BaS_3$ (P-$42_1$m) | -1.231 | a = 6.9836, c = 4.2418, | a = 6.871, 4.168[5] | 2.397 (direct) |
| $ZrS_2$ (P-3m1) | -1.734 | a = 3.691, c = 6.611, | a = 3.663, c = 5.821[6] | 1.595 (indirect) |
| $ZrS_3$ ($P2_1$/m) | -1.378 | a = 3.6572, b = 5.2023, c = 9.504, | a = 3.624, b = 5.124, c = 8.98[7] | 1.876 (indirect) |
| ZrS (Fm-3m) | -1.459 | a = 5.24814, | a = 5.240[8] | 0 |
| ZrS (P4/nmm) | -1.592 | a = 3.61935, c = 5.54735, | a = 3.55, c = 6.31[9] | 0 |
| $Zr_3S_4$ (Fd-3m) | -1.371 | a = 10.37, | a = 10.25[8] | 0 |
| BaZrS$_3$ (Pnma) | -2.071 | a = 6.922, b = 7.0671, c= 9.9063, | a = 7.025, b = 7.06, c = 9.9813[10] | 1.724 (direct) |
| Ba$_2$ZrS$_4$ (I4/mmm) | -2.163 | a = 5.011, c = 15.8994, | a = 4.785 , c = 15.964[11] | 1.289 (indirect) |
| Ba$_3$Zr$_2$S$_7$ (P4$_2$/mnm) | -2.127 | a = 7.1694, c = 25.5525, | a = 7.709 c = 25.437[12] | 1.542 (indirect) |
| Ba$_3$Zr$_2$S$_7$ (I4/mmm) | -2.117 | a = 5.043, c = 25.8576, | a = 4.998, c = 25.502[13] | 1.164 (indirect) |
| Ba$_3$Zr$_2$S$_7$ (Cmmm) | -2.118 | a = 7.1381, b = 25.8383, c = 7.1227, | a = 7.0697 b = 25.4923 , c = 7.0269[11] | 1.171 (indirect) |
| Ba$_4$Zr$_3$S$_{10}$ (Fmmm) | | a = 7.0314 , b = 7.0552, c = 35.544 , | a = 7.031, b = 7.0552, c = 35.544[14] | |
| Ba$_4$Zr$_3$S$_{10}$ (I4/mmm) | -2.096 | a = 5.0483, c = 35.9174 | | 1.129 (indirect) |

*Table S2: Phonon calculation details. Supercells and k-grid used for finite difference phonon analysis using Phonopy for every material are collected below. For all structures, the k-density was chosen to be a minimum of 2 points per Å$^{-1}$. Further calculation details are included in the main text.*

| Material (Space Group) | Supercell size | k-grid |
|---|---|---|
| BaS (Fm-3m) | 2x2x2 | 8x8x8 |
| BaS$_2$ (C2/c) | 3x3x2 | 5x5x4 |
| BaS$_3$ (P-42$_1$m) | 3x3x3 | 4x4x5 |
| ZrS$_2$ (P-3m1) | 3x3x3 | 5x5x4 |
| ZrS$_3$ (P2$_1$/m) | 2x3x3 | 4x5x3 |
| ZrS (Fm-3m) | 3x3x3 | 5x5x5 |
| ZrS (P4/nmm) | 3x3x3 | 5x5x5 |
| Zr$_3$S$_4$ (Fd-3m) | 2x2x2 | 6x6x6 |
| BaZrS$_3$ (Pnma) | 2x2x2 | 6x4x6 |
| Ba$_2$ZrS$_4$ (I4/mmm) | 2x2x1 | 8x8x6 |
| Ba$_3$Zr$_2$S$_7$ (P42/mnm) | 2x2x1 | 6x6x2 |
| Ba$_3$Zr$_2$S$_7$ (I4/mmm) | 4x4x1 | 4x4x6 |
| Ba$_3$Zr$_2$S$_7$ (Cmmm) | 2x1x2 | 6x2x6 |
| Ba$_4$Zr$_3$S$_{10}$ (Fmmm) | 2x2x1 | 6x6x4 |
| Ba$_4$Zr$_3$S$_{10}$ (I4/mmm) | 2x2x1 | 6x6x4 |

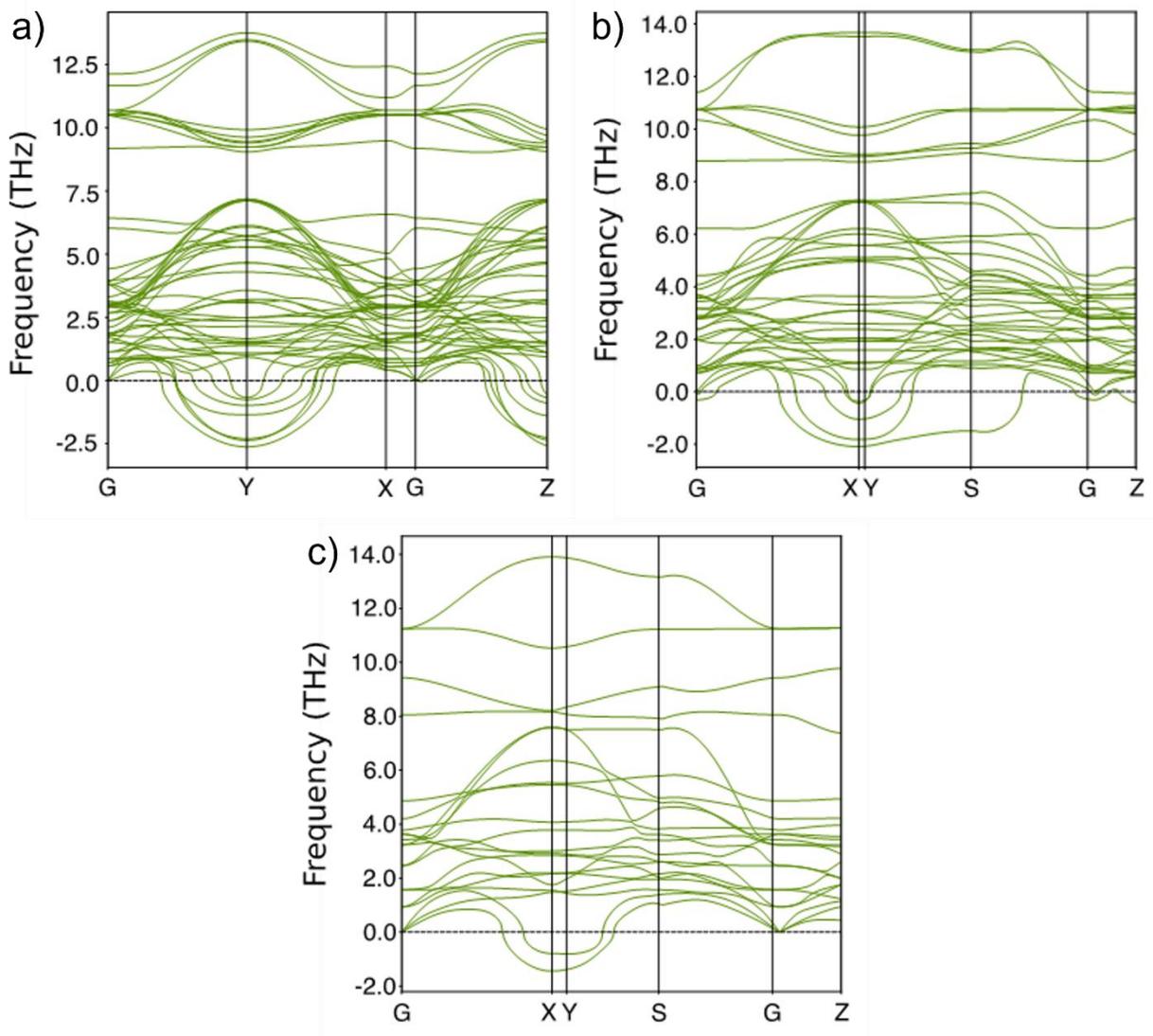

*Figure S4. Phonon dispersions showing imaginary modes for a) $Ba_4Zr_3S_{10}$ in the Fmmm phase b) $Ba_3Zr_2S_7$ in the Immm phase c) $Ba_2ZrS_4$ in the Immm phase. As is common in the literature, we omitted the imaginary modes from our calculation of the thermodynamic partition function and so they do not contribute to the Gibbs free energy. This approximation can be justified when considering that the thermodynamic properties are calculated statistically across all phonon branches, and that there is a relatively small number of imaginary modes across the whole Brillouin Zone. In addition, it has been shown for a number of related systems that renormalising the imaginary phonon modes to an effective real harmonic frequency, one strategy for inclusion within the Gibbs free energy term, has little impact on the calculated free energies.*[15,16]

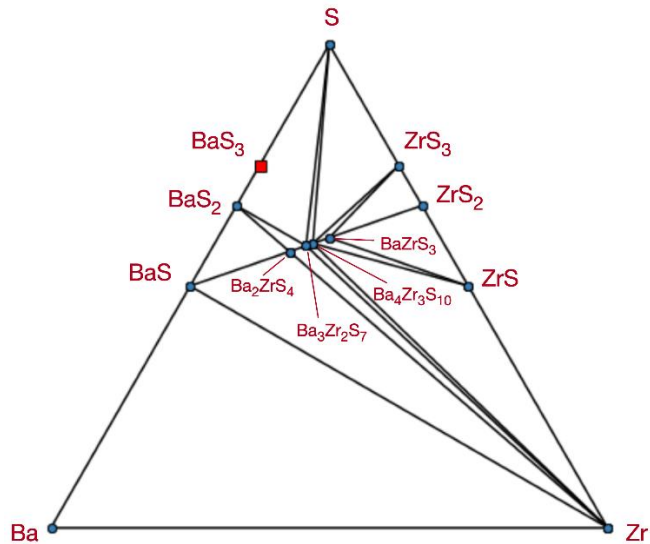

*Figure S2. Phase diagram for the Ba-Zr-S system at 1200K. The phase diagram has been calculated using ThermoPot[17] and Pymatgen.[18] The blue points indicate stable phases on the convex hull, whilst the red square indicates an unstable phase.*

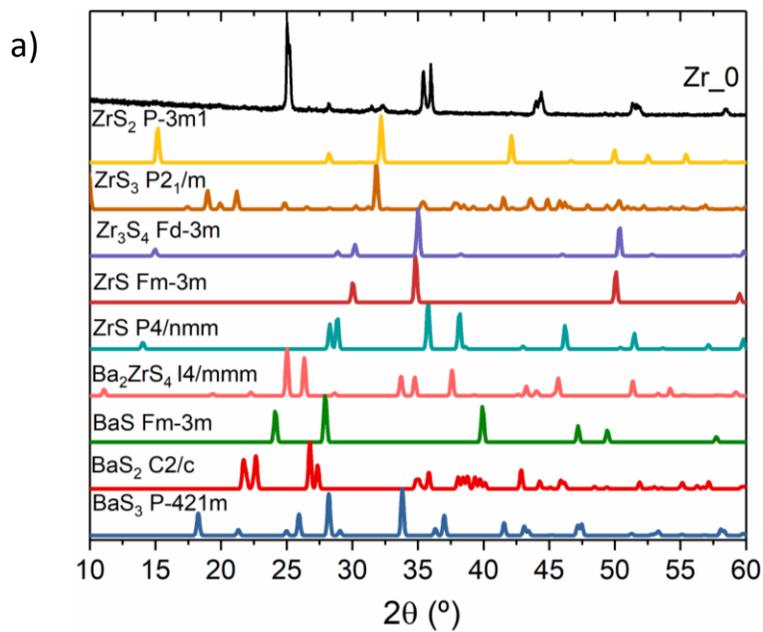

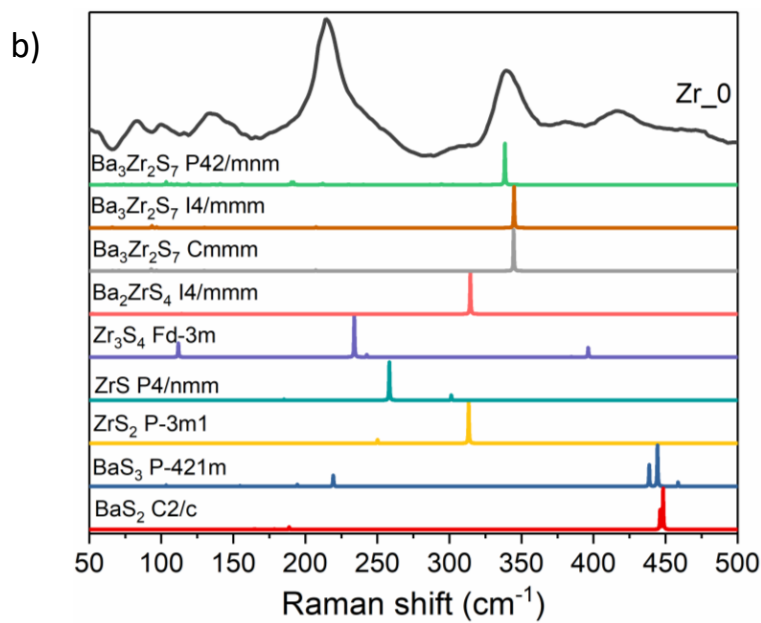

*Figure S3. a) XRD of the binary Zr-S, Ba-S and Ba$_2$ZrS$_4$ phases. b) Raman of the same species compared with the measured sample (the missing Ba-S and Zr-S compositions are not Raman active).*

*Table S3. List of possible undesired reaction by-products. Y indicates matching between the reference spectra/diffractogram and the experimental data, while X indicates no match. In the Raman column the numbers in brackets represent the ICSD. In the XRD column the numbers in brackets represent the ICSD collection code used as reference.*

|  | **Raman** | **XRD** |
|---|---|---|
| **ZrO$_2$ (monoclinic)** | X [19] | Y (18190) |
| **ZrO$_2$ (tetragonal)** | X [20] | X (66781) |
| **ZrOS** | No available data | X (6166) |
| **Zr(SO$_4$)$_2$** | X [21] | X (16001) |
| **ZrC** | X [22-24] | X (159874) |
| **BaO** | X [25] | X (26961) |
| **BaO$_2$** | X [26] | X (72509) |
| **BaCO$_3$** | X [24] | X (15196) |
| **BaSO$_4$** | X [27] | X (16917) |

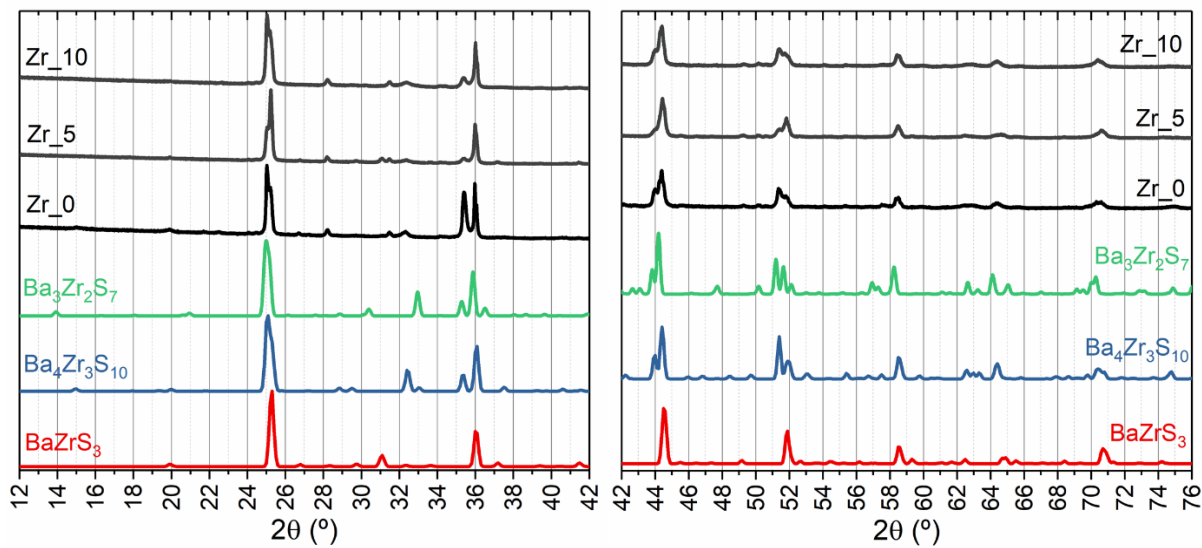

*Figure S4. Left: X-ray diffraction at (left) low angle and (right) high angle for Zr_0, Zr_5 and Zr_10 and the reference diffractograms for $BaZrS_3$, $Ba_4Zr_3S_{10}$ and $Ba_3Zr_2S_7$.*

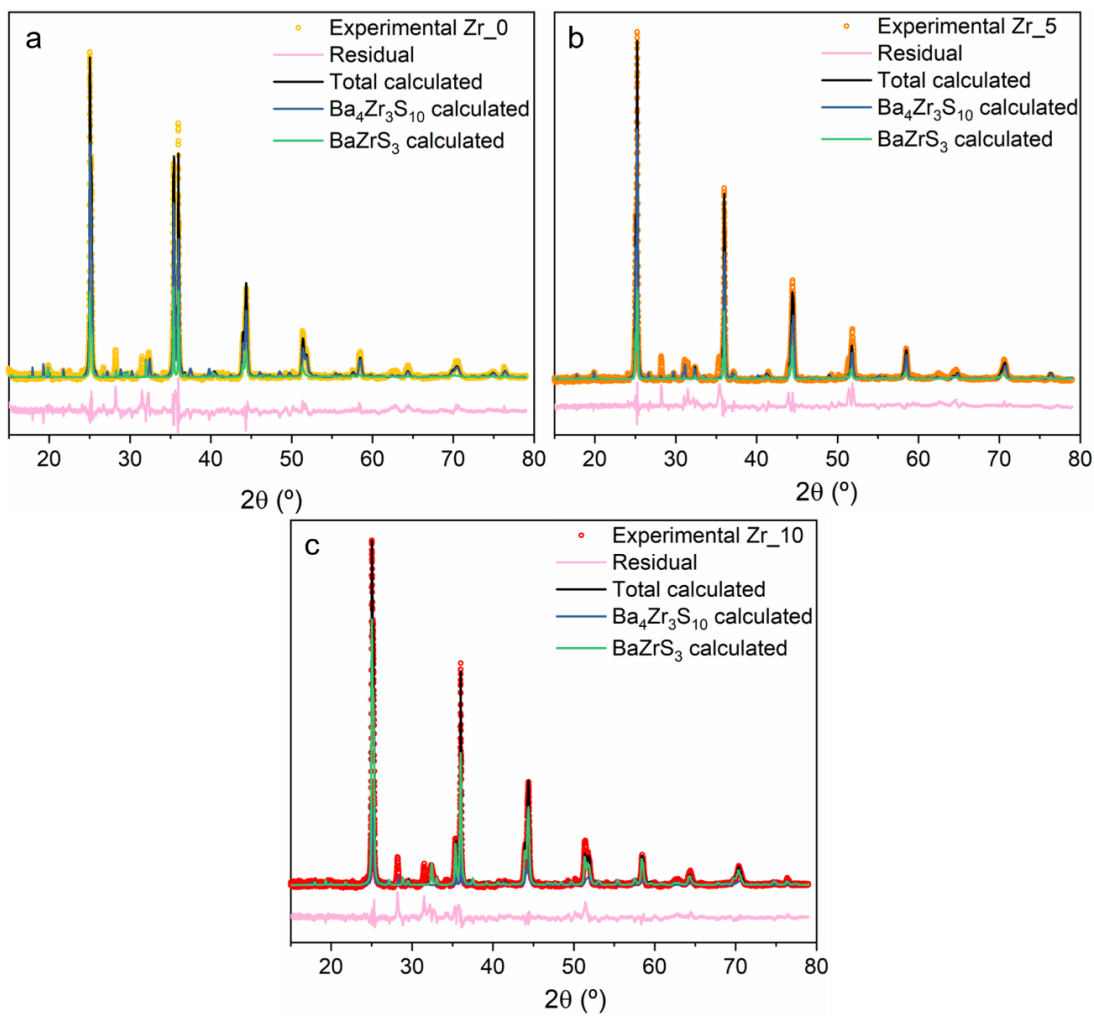

*Figure S5. Rietveld refinement of Zr_0 (a), Zr_5 (b), Zr_10 (c).*

Table S4. Quantifications, lattice parameters and fractional coordinates derived from Rietveld refinement on Zr_0

| **Zr_0** | | \<td colspan=3\>$Ba_{0.74}Zr_{0.84}S_{2.92}$ | | | | \<td colspan=3\>$Ba_{3.52}Zr_{2.83}S_{8.84}$ |
|---|---|---|---|---|---|---|---|---|---|
| | | Pnma (62) | | | | | Fmmm (69) | | |
| | Wt (%) | 31 | | | | | 69 | | |
| | a (Å) | 7.0747 | | | | | 7.0382 | | |
| | b (Å) | 10.1537 | | | | | 7.0584 | | |
| | c (Å) | 7.0446 | | | | | 35.4219 | | |
| | | | | | | | | | |
| Site | | x | y | z | Site | | x | y | z |
| 4c | Ba | 0.0358 | 0.2500 | -0.0197 | 8i | Ba | 0.5000 | 0.0000 | 0.0694 |
| 4b | Zr | 0.0000 | 0.0000 | 0.5000 | 8i | Ba | 0.5000 | 0.0000 | 0.2087 |
| 4c | S | 0.0027 | 0.2500 | 0.5017 | 4a | Zr | 0.0000 | 0.0000 | 0.0000 |
| 8d | S | 0.1874 | 0.0001 | 0.8270 | 8i | Zr | 0.0000 | 0.0000 | 0.1447 |
| | | | | | 8i | S | 0.0000 | 0.0000 | 0.0675 |
| | $R_{wp}$ | 16.03 % | | | 8i | S | 0.0000 | 0.0000 | 0.2093 |
| | $R_{exp}$ | 10.1% | | | 8e | S | 0.2500 | 0.2500 | 0.0000 |
| | | | | | 16j | S | 0.2500 | 0.2500 | 0.1323 |

Table S5. Quantifications, lattice parameters and fractional coordinates derived from Rietveld refinement on Zr_5

| **Zr_5** | | $Ba_{0.79}Zr_{0.83}S_{2.76}$ | | | | | $Ba_{3.71}Zr_{2.58}S_{8.47}$ | | |
|---|---|---|---|---|---|---|---|---|---|
| | | Pnma (62) | | | | | Fmmm (69) | | |
| | Wt (%) | 59 | | | | | 41 | | |
| | a (Å) | 7.1119 | | | | | 7.0686 | | |
| | b (Å) | 9.9736 | | | | | 7.1512 | | |
| | c (Å) | 7.0345 | | | | | 34.9998 | | |
| | | | | | | | | | |
| Site | | x | y | z | Site | | x | y | z |
| 4c | Ba | 0.0362 | 0.2500 | 0.0000 | 8i | Ba | 0.5000 | 0.0000 | 0.0703 |
| 4b | Zr | 0.0000 | 0.0000 | 0.5000 | 8i | Ba | 0.5000 | 0.0000 | 0.2226 |
| 4c | S | 0.0019 | 0.2500 | 0.5962 | 4a | Zr | 0.0000 | 0.0000 | 0.0000 |
| 8d | S | 0.1874 | 0.0042 | 0.7236 | 8i | Zr | 0.0000 | 0.0000 | 0.1429 |
| | | | | | 8i | S | 0.0000 | 0.0000 | 0.0638 |
| | $R_{wp}$ | 14.13 % | | | 8i | S | 0.0000 | 0.0000 | 0.2029 |
| | $R_{exp}$ | 9.3% | | | 8e | S | 0.2500 | 0.2500 | 0.0000 |
| | | | | | 16j | S | 0.2500 | 0.2500 | 0.1421 |

*Table S6. Quantifications, lattice parameters and fractional coordinates derived from Rietveld refinement on Zr_10*

| **Zr_10** | | $Ba_{0.77}Zr_{0.83}S_{2.68}$ | | | | | $Ba_{3.88}Zr_{2.52}S_{8.52}$ | | |
| --- | --- | --- | --- | --- | --- | --- | --- | --- | --- |
| | | **Pnma (62)** | | | | | **Fmmm (69)** | | |
| | *Wt (%)* | 14 | | | | | 86 | | |
| | **a (Å)** | 7.0214 | | | | | 7.0392 | | |
| | **b (Å)** | 10.0906 | | | | | 7.0674 | | |
| | **c (Å)** | 7.041 | | | | | 35.5196 | | |
| | | | | | | | | | |
| *Site* | | *x* | *y* | *z* | *Site* | | *x* | *y* | *z* |
| *4c* | Ba | 0.0310 | 0.2500 | -0.0097 | *8i* | Ba | 0.5000 | 0.0000 | 0.0694 |
| *4b* | Zr | 0.0000 | 0.0000 | 0.5000 | *8i* | Ba | 0.5000 | 0.0000 | 0.2087 |
| *4c* | S | 0.0002 | 0.2500 | 0.5475 | *4a* | Zr | 0.0000 | 0.0000 | 0.0000 |
| *8d* | S | 0.1898 | 0.0092 | 0.7822 | *8i* | Zr | 0.0000 | 0.0000 | 0.1447 |
| | | | | | *8i* | S | 0.0000 | 0.0000 | 0.0675 |
| | **R$_{wp}$** | *12.87 %* | | | *8i* | S | 0.0000 | 0.0000 | 0.2093 |
| | **R$_{exp}$** | *9.21%* | | | *8e* | S | 0.2500 | 0.2500 | 0.0000 |
| | | | | | *16j* | S | 0.2500 | 0.2500 | 0.1323 |

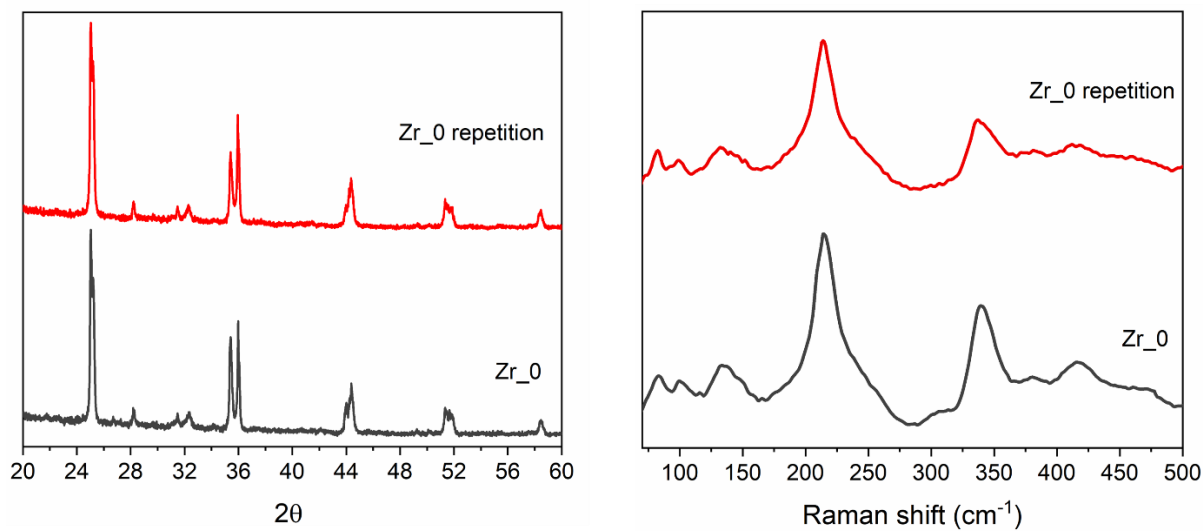

*Figure S6. Left: XRD of the Zr_0 and of another synthetic batch produced with the same Zr_0 conditions. Right: Raman spectra of the Zr_0 and of another synthetic batch produced with the same Zr_0 conditions.*

*Table S7: Predicted decomposition products at various Ba-Zr-S stoichiometries. Each row corresponds to a starting composition within the error range of measured EDS values. The three right hand columns each correspond to a degradation product, with the values within each column corresponding mole percentages.*

|  | Starting composition | Products | | |
|---|---|---|---|---|
|  |  | $BaZrS_3$ | $Ba_4Zr_3S_{10}$ | $ZrS$ |
| Zr_0 | $Ba_{0.21}Zr_{0.19}S_{0.59}$ | 0.61 | 0.39 | 0.01 |
| Zr_5 | $Ba_{0.20}Zr_{0.21}S_{0.58}$ | 0.34 | 0.56 | 0.09 |
| Zr_10 | $Ba_{0.21}Zr_{0.20}S_{0.57}$ | 0.11 | 0.81 | 0.08 |

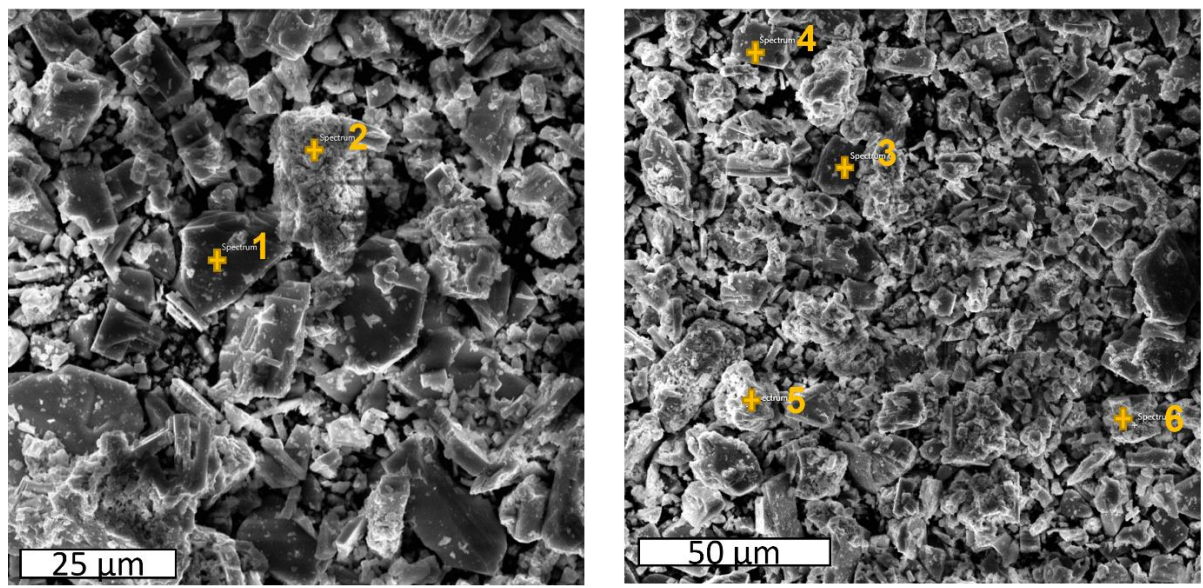

| Zr_0 | 1 | 2 | 3 | 4 | 5 | 6 |
|---|---|---|---|---|---|---|
| Ba | 23 | 19 | 25 | 22 | 17 | 28 |
| Zr | 20 | 48 | 20 | 21 | 28 | 20 |
| S | 57 | 33 | 55 | 57 | 55 | 52 |

*Figure S7: Analysis of the composition variation of the sample Zr_0. The different points analysed are indicated in the pictures with a cross, and the corresponding composition is reported in the table below the figure.*

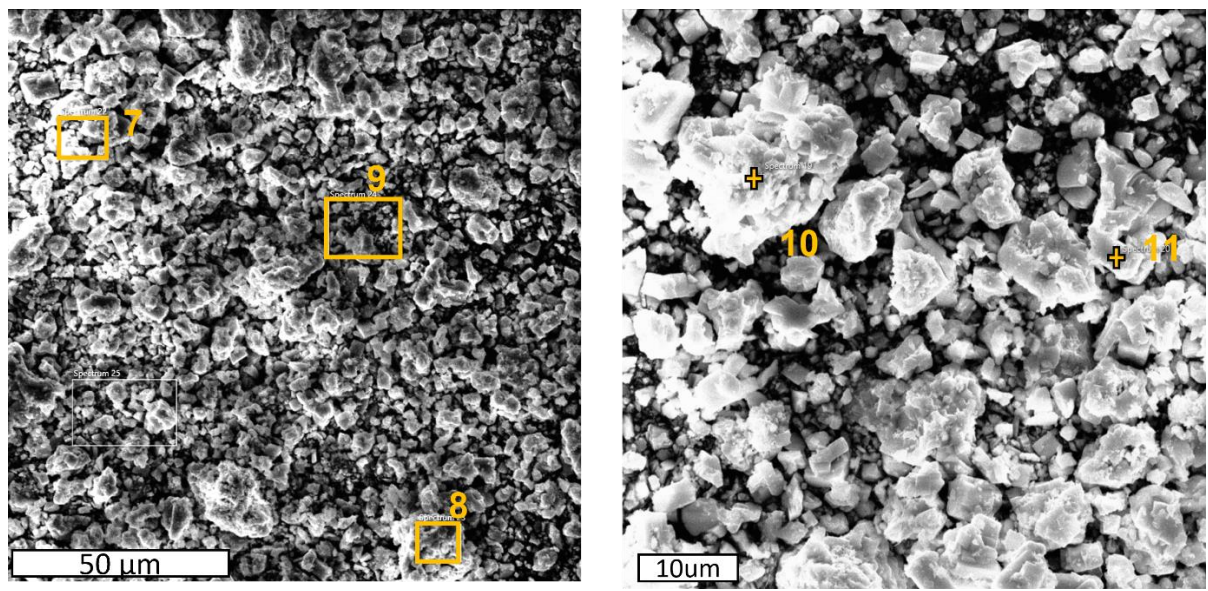

| Zr_5 | 7 | 8 | 9 | 10 | 11 |
|---|---|---|---|---|---|
| Ba | 24 | 24 | 24 | 28 | 21 |
| Zr | 21 | 22 | 25 | 21 | 21 |
| S | 55 | 54 | 51 | 51 | 58 |

*Figure S8. Analysis of the composition variation of the sample Zr_5. The different points analysed are indicated in the pictures with a cross or a rectangle, and the corresponding composition is reported in the table below the figure.*

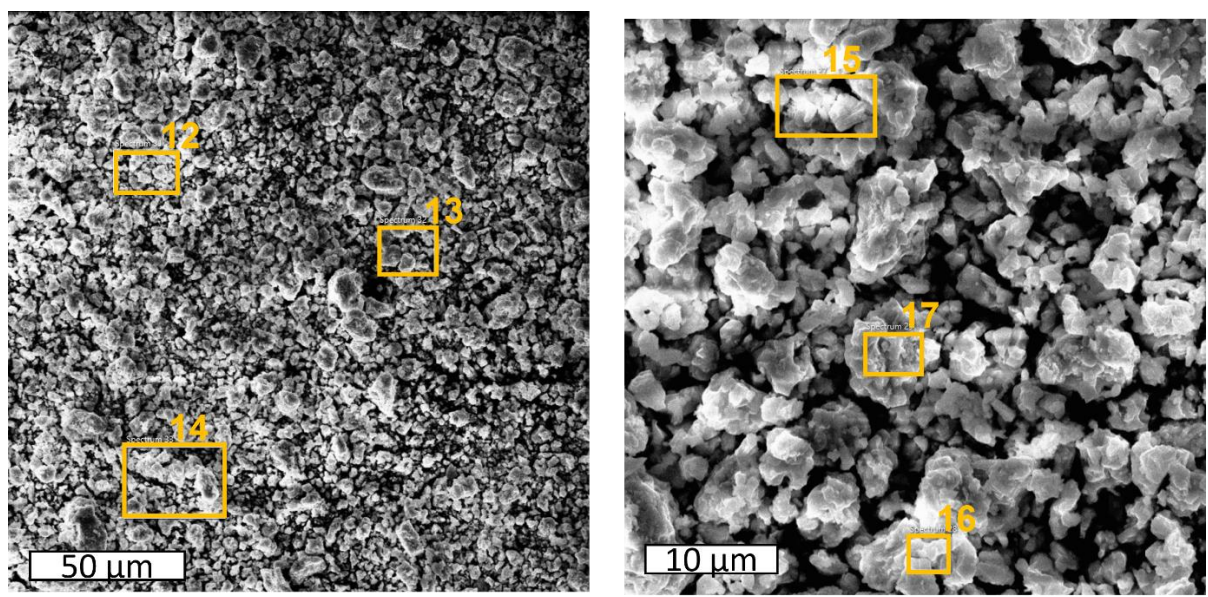

| Zr_10 | 12 | 13 | 14 | 15 | 16 | 17 |
|---|---|---|---|---|---|---|
| Ba | 25 | 25 | 24 | 24 | 25 | 24 |
| Zr | 22 | 22 | 23 | 23 | 21 | 23 |
| S | 53 | 53 | 53 | 53 | 54 | 53 |

*Figure S9. Analysis of the composition variation of the sample Zr_10. The different points analysed are indicated in the pictures with a cross or a rectangle, and the corresponding composition is reported in the table below the figure.*

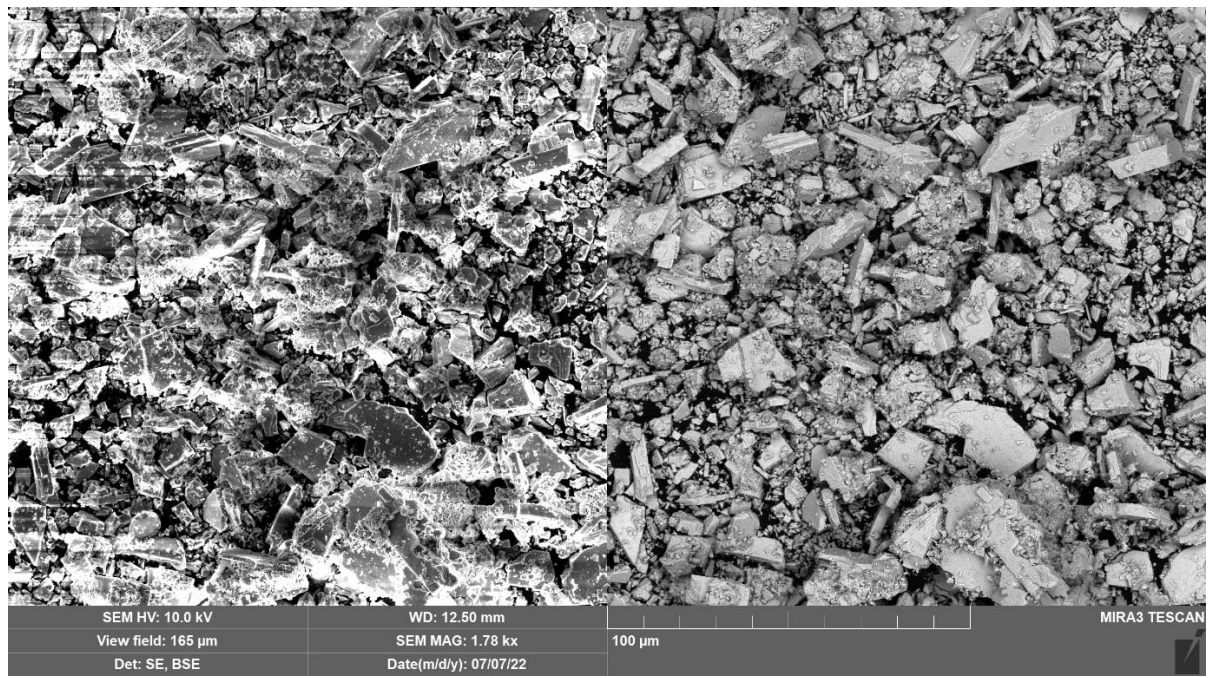

*Figure S10. Sem image of the Zr_0 sample. Left: Image obtained from the secondary electron detection. Right: Image obtained from the back-scattered electron detection*